\documentclass[11pt, conference]{IEEEtran}
\IEEEoverridecommandlockouts
\usepackage{cite}
\usepackage{amsmath,amssymb,amsfonts}
\usepackage{graphicx}
\usepackage{textcomp}
\usepackage{xcolor}
\usepackage{graphicx,color}
\usepackage{algorithm}
\usepackage{booktabs}
\usepackage{multirow}
\usepackage{tabularx,lipsum}
\usepackage{float}
\def\BibTeX{{\rm B\kern-.05em{\sc i\kern-.025em b}\kern-.08em
    T\kern-.1667em\lower.7ex\hbox{E}\kern-.125emX}}

\begin{document}
\renewcommand{\L}{\mathcal{L}}

\title{\bf \LARGE PAD-FT: A Lightweight Defense for Backdoor Attacks via Data Purification and
Fine-Tuning
}
\author{\IEEEauthorblockN{Yukai Xu, Yujie Gu, and Kouichi Sakurai \\[0.3cm]}
\IEEEauthorblockA{Faculty of Information Science and Electrical Engineering, 
Kyushu University, Fukuoka, Japan} 
}

\maketitle

\begin{abstract}
Backdoor attacks pose a significant threat to deep neural networks, particularly as recent advancements have led to increasingly subtle implantation, making the defense more challenging. Existing defense mechanisms typically rely on an additional clean dataset as a standard reference and involve retraining an auxiliary model or fine-tuning the entire victim model. However, these approaches are often computationally expensive and not always feasible in practical applications. 
In this paper, we propose a novel and lightweight defense mechanism, termed PAD-FT, that does not require an additional clean dataset and fine-tunes only a very small part of the model to disinfect the victim model. To achieve this, our approach first introduces a simple data purification process to identify and select the most-likely clean data from the poisoned training dataset. The self-purified clean dataset is then used for activation clipping and fine-tuning only the last classification layer of the victim model. By integrating data purification, activation clipping, and classifier fine-tuning, our mechanism PAD-FT demonstrates superior effectiveness across multiple backdoor attack methods and datasets, as confirmed through extensive experimental evaluation.
\end{abstract}


\section{Introduction}
\label{sec:intro}
\textbf{Backdoor attacks.} Deep neural networks (DNNs) have achieved significant success across various domains \cite{gpt3,wav2vec2}, especially in image recognition due to their high effectiveness \cite{imagenet,yolov3}.
However, training DNNs requires large amounts of labeled training data, often sourced from untrusted third parties, which introduces substantial security risks. Among these risks, backdoor attacks pose a critical threat.

BadNets \cite{gu2017badnets}, the first and most representative backdoor attack, randomly selects a subset of samples from the original benign dataset, embeds a backdoor trigger into these benign images and changes their labels to an attacker-specified target label. These poisoned samples are then mixed with the remaining benign samples to create a poisoned training dataset, which is subsequently used for model training.
Blend \cite{Blended} enhanced backdoor attacks by blending benign images with an entire trigger image, making the attack more potent. More recently, even subtler and more effective attacks have been proposed, such as SIG \cite{SIG}, which uses a backdoor signal as the trigger pattern, and WaNet \cite{wanet}, which employs image warping as the backdoor trigger, rendering these attacks even more inconspicuous.
The increasing subtlety of these backdoor attacks significantly raises the difficulty of detection and prevention in practical scenarios.

\textbf{Backdoor defense.} 
In the literature, many defense mechanisms have been proposed to defend against or mitigate backdoor attacks, which can be broadly categorized into two types: in-training defense and post-training defense. In-training defense methods are applied during the training process, assuming the defender is aware of the existence of the attack. DBD \cite{DBD} is a representative in-training defense mechanism that decouples the training process into three stages: self-supervised pre-training of the feature extractor, supervised learning of the classifier, and semi-supervised learning of the entire model. However, this approach involves a complex training process, increasing both time and computational costs.




Post-training defense methods focus on disinfecting models that have already been compromised by backdoor attacks. For example, \cite{nad} utilizes a teacher model pre-trained on a clean dataset to perform knowledge distillation on the victim student model. However, this approach involves an additional model and requires an external clean dataset, which is often impractical in real-world scenarios.
\cite{MM-BD} proposes a post-training backdoor detection method that leverages the maximum margin of activation values. In this approach, the defender is assumed to have access to a small clean dataset, which is used to detect if the model is poisoned and to optimize an activation clipping upper bound that reduces the activation margin, thereby disinfecting the victim model.
However, obtaining a reliable clean dataset is not always feasible in practical applications.


\textbf{Our approach and contributions.} 
To address the issues in existing defense methods that require additional clean dataset and incur high computational costs, we propose a lightweight post-training defense method, termed PAD-FT.
This method does not require additional clean data and fine-tunes only a small portion of the victim DNN model.

More precisely, PAD-FT first leverages a simple data purification process that employs the symmetric cross-entropy (SCE) \cite{sce} as a metric to identify and select the most-likely clean data from the poisoned dataset, thereby creating a self-purified clean dataset without introducing external data. Next, PAD-FT applies an activation clipping process using optimized clipping bounds derived from the self-purified clean dataset. Finally, PAD-FT fine-tunes only the classifier with the activation clipping to enhance the robustness of the defense while reducing computational costs. 
Extensive experiments demonstrate the effectiveness and superiority of the proposed PAD-FT in defending against backdoor attacks.


To sum up, our contributions in this paper are as follows. 
\begin{itemize}
    \item We propose an easy-to-implement data purification approach to select the most-likely clean data from the poisoned dataset, thereby creating a self-purified clean dataset without introducing external data, making it more practical for real-world applications.\\[-0.4cm]
    \item  
    We propose a novel and lightweight backdoor defense mechanism, PAD-FT, by integrating data purification, activation clipping, and classifier fine-tuning, avoiding the use of additional models or data and demonstrating very low computational cost.\\[-0.4cm]
    \item 
    We conduct comprehensive experimental evaluations on the proposed mechanism PAD-FT, demonstrating its effectiveness and superiority against a variety of backdoor attack strategies across diverse datasets.

\end{itemize}

\section{Preliminary}
\label{sec:formulation}



Let $F(\cdot)$ denote the image classification model, which consists of feature extractor layers $f_l(\cdot)$, activation layers $a_l(\cdot)$ and a fully connected classifier $\phi(\cdot)$, where $l=0,\dots,L$. 

Let $\mathcal{D} =\{(\mathbf{x}_i, y_i)\}_{i=1}^N$ represent the original dataset, where $\mathbf{x}_i\in\mathcal{X}=[255]^{W\times H\times C}$ is an image sample with width $W$, height $H$ and $C$ channels, and $y_i\in\mathcal{Y}=\{0, 1, \dots
, K\}$ is the corresponding label and $K$ is the number of classes. The output of the model with respect to an input image $\mathbf{x}$ is represented as $$F(\mathbf{x})=\phi\circ a_L \circ f_L \circ \cdots \circ a_0 \circ f_0(\mathbf{x}).$$

A typical backdoor attack mechanism embeds a specific pattern $\mathbf{p}$ into an original sample $\mathbf{x}$, generating a poisoned dataset $\mathcal{D}_p = \{(\mathbf{x'}_i, y'_i)\}_{i=1}^{N_p}$, where $\mathbf{x'}=\mathbf{x}+\mathbf{p}$ and $y'$ is the target class. 
The remaining data remains benign and forms the benign subset $\mathcal{D}_b$.The final training dataset is then $$\mathcal{D}_t = \mathcal{D}_b \cup \mathcal{D}_p.$$
The poison rate $\rho$ is calculated as $\rho=\frac{N_p}{N}$, where $N_p$ is the number of poisoned samples and $N$ is the total number of samples in the dataset. As the amount of poisoned data increases, the poisoning rate $\rho$ increases accordingly.

\section{Method}
\label{sec:backdoor}

Our method PAD-FT consists of three key components: data purification, activation clipping, and classifier fine-tuning, which are described below. The entire framework of PAD-FT is outlined in Algorithm \ref{alg:def}. 

\subsection{Data Purification}
\label{subsec:data_purf}
To avoid resorting to external data, we aim to design a simple data purification method to identify and select the most-likely clean data from the poisoned training dataset. To that end, we employ symmetric cross-entropy (SCE) loss \cite{sce} as an evaluation metric for data purification, which differs from the model's training loss function. SCE combines the traditional cross-entropy loss with a reverse cross-entropy term, enhancing the model's robustness and helping to filter out the most-likely clean data from noisy data. 

For each data point $(\mathbf{x}_i,y_i)\in\mathcal{D}_t$ in the poisoned training dataset, we calculate the SCE loss $\ell_{SCE}$ as
\begin{equation}\label{eq:sce}
    \begin{aligned}
    \ell_{SCE}(\mathbf{x}_i, y_i;F) = &\ \alpha\cdot\ell_{CE}(\sigma(F(\mathbf{x}_i)), y_i) \\
    &+ (1-\alpha)\cdot\ell_{CE}(y_i, \sigma(F(\mathbf{x}_i)))
\end{aligned}
\end{equation}
where $\ell_{CE}$ is the standard cross-entropy loss, $\sigma(\cdot)$ is the softmax function, $F$ is the classification model, and $\alpha$ is a hyperparameter to balance the two terms.

After calculating $\ell_{SCE}$ for all training data in the dataset, we select $n_c$ images from each class that have the smallest SCE loss values. These selected images are regarded as pseudo-clean data, as their low loss values indicate that the model is confident in their correct classification, even in the presence of poisoned data. The selected pseudo-clean dataset is denoted as $\mathcal{D}_c$, containing $N_c=n_c\times K$ images. 
A smaller $N_c$ indicates that the purified dataset $\mathcal{D}_c$ contains clean data with greater confidence.

\subsection{Activation Clipping}


As highlighted in \cite{gu2017badnets}, backdoor attacks significantly impact activation values. When the victim model is ``activated" by the trigger pattern, the associated activation nodes produce abnormally high outputs, leading to incorrect classification results. To mitigate this, \cite{MM-BD} proposed setting an upper bound on the activation layers using an additional clean dataset. 
This strategy clips the abnormally high activation values triggered by the pattern to normal levels, using an external clean dataset as a reference standard.


Considering that an external clean dataset is usually not feasible in real-world applications, our method PAD-FT integrates a self-purified data subset obtained in Section \ref{subsec:data_purf} into the activation clipping strategy. Its effectiveness in defending the victim model is demonstrated in Section \ref{sec:experiment}.

Specifically, for the $l$-th activation layer $a_l(\cdot)$, an upper bound $\mathbf{z}_l$ is introduced to clip the activation output, where the clipped activation is represented by $\Bar{a}_l(\cdot)=\min(a_l(\cdot),\mathbf{z}_l)$ and  the corresponding bounded logits of the model are denoted as $\Bar{F}(\cdot)$. 
Let $\mathbf{Z}=\{\mathbf{z}_0, \dots, \mathbf{z}_L\}$ represent the set of clipping bounds for each activation layer. 
As in \cite{MM-BD}, the upper bounds are optimized using the following loss function
\begin{equation}\label{eq:activtionclip}
\begin{aligned}
    \L_{AC}
    =& \frac{1}{N_c}\sum_{\mathbf{x}\in \mathcal{D}_c}\ell_{MSE}(\Bar{F}(\mathbf{x};\mathbf{Z}),F(\mathbf{x})) 
    + \lambda\sum_l||\mathbf{z}_l||_2
\end{aligned}
\end{equation}
where $\ell_{MSE}$ is the mean squared error loss and $\lambda$ is dynamically adjusted as in \cite{MM-BD}. 
By minimizing this loss function on the selected purified clean dataset $\mathcal{D}_c$, the clipping bounds for activation values can be established. Accordingly, the victim model can be disinfected by using these bounds to clip abnormally large activation values to normal levels.


\subsection{Classifier Fine-tuning}

After activation clipping, we employ fine-tuning to enhance model performance. In existing methods, the fine-tuning process for backdoor defenses typically requires a clean dataset and involves updating the entire model, e.g. \cite{ftsam}, which is computationally expensive, especially for large models.

In contrast, our method PAD-FT employs a self-purified clean dataset and fine-tunes only the classifier, thereby significantly reducing computational cost. Experimental results indicate the effectiveness of PAD-FT, as shown in Section \ref{sec:experiment}.

First, inspired by semi-supervised learning, we introduce consistency regularization \cite{consistencyregularization} to enhance the robustness of our backdoor defense.
The consistency regularization loss is
\begin{equation}
        \L_{CR}=\frac{1}{N_c}\sum_{\mathbf{x} \in \mathcal{D}_c}\ell_{CE}\left(\sigma(\Bar{F}(\gamma(\mathbf{x}))), \sigma(\Bar{F}(\mathbf{x}))\right)
\end{equation}
where $\gamma(x)$ represents an augmented version of the image $x$ via techniques such as  
flipping, brightness adjustments, and contrast modification, and 
$\Bar{F}(\cdot)$ is the clipped model.
Consistency regularization encourages the classifier to make consistent predictions on both the original and augmented images, thereby enhancing robustness against backdoor attacks.

Then the loss function used for classifier fine-tuning is
\begin{equation}\label{eq:ft}
        \L_{FT} = \beta\cdot\frac{1}{N_c}\sum_{\mathbf{x}\in\mathcal{D}_c}\ell_{SCE}(\mathbf{x},y;\Bar{F}) + (1-\beta)\cdot\L_{CR}
\end{equation}
where $\beta$ 
is a hyperparameter that balances the contributions of the SCE loss and the consistency regularization loss.

The overall process of the proposed backdoor defense mechanism PAD-FT is illustrated in Algorithm ~\ref{alg:def}.

\section{Experiments}
\label{sec:experiment}

\begin{table*}[h!]
    \centering
    \caption{ACC (\%) and ASR (\%) of defense mechanisms with poison rate $\rho=10\%$. ($\uparrow$ indicates that a higher value is better, while $\downarrow$ indicates that a lower value is preferred.)}
    \begin{tabular}{c c  c c  c c c c c c c c}
    \toprule 
    \multirow{2}{*}{Dataset} &\multirow{2}{*}{Attack} & \multicolumn{2}{c}{No Defense} & \multicolumn{2}{c}{DBD}  & \multicolumn{2}{c}{MM-BD} &\multicolumn{2}{c}{\textbf{PAD}} & \multicolumn{2}{c}{\textbf{PAD-FT}}\\
     \cmidrule(lr){3-4} \cmidrule(lr){5-6} \cmidrule(lr){7-8} \cmidrule(lr){9-10} \cmidrule(lr){11-12} 
    && ACC\,$\uparrow$ & ASR\,$\downarrow$ & ACC\,$\uparrow$ & ASR\,$\downarrow$ & ACC\,$\uparrow$ & ASR\,$\downarrow$ & ACC\,$\uparrow$ & ASR\,$\downarrow$ & ACC\,$\uparrow$ & ASR\,$\downarrow$\\
\midrule
 \multirow{3}{*}{CIFAR-10}& BadNets & 91.82 & 93.79 & 79.89 & 6.83  & 87.33 & 28.07 & 83.28 & 28.13 & 85.34 & 6.62 \\
 & Blended & 93.69 & 99.76 & 90.11 & 7.32  & 84.57 & 29.11 & 82.97 & 28.47 & 83.40 & 6.56 \\
 & WaNet & 90.57 & 96.93 & 85.53 & 9.17  & 84.38 & 20.20 & 84.84 & 20.18 & 86.86 & 8.92 \\
 \midrule
\multirow{3}{*}{CIFAR-100} & BadNets & 67.19 & 85.95 & 54.56 & 91.48  & 59.88 & 0.15 & 62.03 & 0.21 & 57.03 & 0.14 \\
 & Blended & 69.43 & 99.44 & 57.61 & 99.90 & 60.05 & 0.45 & 62.90 & 0.44 & 50.56 & 0.29\\
 & WaNet & 71.10 & 100.00 & 55.84 & 97.19  & 58.30 & 0.70 & 62.36 & 0.69 & 54.94 & 0.47\\
    \bottomrule
    \end{tabular}
    \label{tab:results10}
\end{table*}

\begin{table*}[h!]
    \centering
    \caption{ACC (\%) and ASR (\%) of defense mechanisms with poison rate $\rho=5\%$}
    \begin{tabular}{c c  c c  c c c c c c c c}
    \toprule 
    \multirow{2}{*}{Dataset} &\multirow{2}{*}{Attack} & \multicolumn{2}{c}{No Defense} & \multicolumn{2}{c}{DBD} & \multicolumn{2}{c}{MM-BD} &\multicolumn{2}{c}{\textbf{PAD} } &\multicolumn{2}{c}{\textbf{PAD-FT}}\\
    \cmidrule(lr){3-4} \cmidrule(lr){5-6} \cmidrule(lr){7-8} \cmidrule(lr){9-10} \cmidrule(lr){11-12}
    && ACC\,$\uparrow$ & ASR\,$\downarrow$ & ACC\,$\uparrow$ & ASR\,$\downarrow$ & ACC\,$\uparrow$ & ASR\,$\downarrow$ & ACC\,$\uparrow$ & ASR\,$\downarrow$ & ACC\,$\uparrow$ & ASR\,$\downarrow$\\
\midrule
 \multirow{3}{*}{CIFAR-10}& BadNets & 92.35 & 89.52 &68.87 & 8.86  & 85.97 & 11.33 & 85.73 & 12.40 & 82.46 & 8.36 \\
 & Blended & 93.76 & 99.31 & 67.44 & 99.96  & 85.13 & 10.09 & 85.09 & 9.55 & 81.09 & 7.44 \\
 & WaNet & 91.45 & 99.98 & 86.97 & 4.95  & 85.75 & 13.27 & 86.36 & 13.11 & 86.03 &10.50 \\
 \midrule
\multirow{3}{*}{CIFAR-100} & BadNets & 69.65 & 69.32 & 62.25 & 0.30 & 62.99 & 0.21 & 66.84 & 0.17 & 60.08 & 0.12 \\
 & Blended & 70.32 & 98.72 & 59.53 & 99.89  & 63.79 & 0.05 & 67.36 & 0.01  &  63.00 & 0.12\\
 & WaNet & 71.94 & 99.94 & 61.80 & 1.26  & 62.93 & 0.39 & 65.93 & 0.12 & 62.78 & 0.07 \\
    \bottomrule
    \end{tabular}
    \label{tab:results5}
\end{table*}

\subsection{Settings}
We conduct experimental evaluations of the backdoor defense mechanisms with the following settings.

\begin{algorithm}[H]
\caption{The Proposed PAD-FT Defense Mechanism}\label{alg:def}
\textbf{Input: }Poisoned training dataset $\mathcal{D}_t$, the victim model $F(\cdot)$, the amount $N_c$ of purified dataset \\
\textbf{Step 1: Data Purification} \\
    \text{1: } Calculate the SCE loss for each image in $\mathcal{D}_t$ via \eqref{eq:sce}. \\
    \text{2: } Select the images with the top-$n_c$ smallest loss values\\ \text{\quad} from each class to form a purified clean dataset $\mathcal{D}_c$.  \\
\textbf{Step 2: Activation Clipping}\\
    \text{3:\ } Initialize the clipping bounds $\mathbf{Z}$ with very large \\ \text{\quad} values.\\
    \text{4:\ } Update $\mathbf{Z}$ by gradient descent via \eqref{eq:activtionclip}.\\
\textbf{Step 3: Classifier Fine-tuning}\\
    \text{5:\ } Generate augmented images $\gamma(\mathbf{x})$ for each $\mathbf{x}\in \mathcal{D}_c$.\\
    \text{6: } Calculate the fine-tuning loss via \eqref{eq:ft}.\\
    \text{7:\ } Update classifier $\phi(\cdot)$ by gradient descent.

\textbf{Return} the disinfected model $\Bar{F}(\cdot)$
\end{algorithm}

\textbf{Dataset and model.} We utilize two standard datasets: CIFAR-10 and CIFAR-100 \cite{cifar100}. CIFAR-10 consists of 50,000 RGB training images and 10,000 testing images across 10 classes, while CIFAR-100 across 100 classes. We adopt the pre-act ResNet-18 \cite{preactres} as the model architecture in all experiments.

\textbf{Backdoor attack.} To evaluate the performance of the proposed backdoor defense mechanism, we implement three different backdoor attack strategies: BadNets \cite{gu2017badnets}, Blended \cite{Blended} and WaNet \cite{wanet}. 
BadNets uses the most conspicuous backdoor pattern while Blended introduces a more subtle approach, and WaNet represents the most inconspicuous attack strategy.
The poison rate $\rho$ is set as $5\%$ and $10\%$ to show the defense performance under different amount of poisoned data.

\textbf{Metrics.}
We evaluate the performance of the defense mechanisms using two representative metrics:  classification accuracy (ACC) on a clean test dataset, and attack success rate (ASR) on a poisoned test dataset where all samples contain the implanted trigger pattern. 
The adversary's objective is to achieve both high ACC and high ASR, whereas the goal of the defense mechanism is to maintain high ACC while minimizing ASR as much as possible.


\textbf{Implementation.}
For a fair comparison, we follow the default training and defense settings as in BackdoorBench \cite{backdoorbench}, including trigger patterns, learning rates, weight decay, and other relevant hyperparameters. We compare the proposed defense mechanism PAD-FT, as well as its preliminary stage before fine-tuning (denoted as PAD), with state-of-the-art defense mechanism baselines: DBD \cite{DBD}, MM-BD \cite{MM-BD}.

For the proposed PAD-FT method, we adopt $\alpha=0.5$ as in \cite{sce} and $\beta=0.5$  to achieve a balance between the SCE loss and the consistency regularization loss in \eqref{eq:ft}. We adopt $N_c=2500$ on both datasets during the experiments.

\subsection{Results}
Table \ref{tab:results10} and Table \ref{tab:results5} present the empirical results of our PAD-FT, PAD and other baselines across various datasets and backdoor attack scenarios. 
The results demonstrate that both PAD-FT and PAD offer significant advantages over other defense mechanisms. 

Notably, PAD and PAD-FT maintain a strong balance between ACC and ASR. 
For instance, in CIFAR-10 (BadNets at $\rho=5\%$), PAD-FT achieves an ACC of $82.46\%$ and an ASR of $8.36\%$, while DBD sacrifices accuracy, yielding only $68.87\%$ ACC for a similar ASR. 

Additionally, PAD-FT’s fine-tuning mechanism significantly improves ASR, as seen in CIFAR-10, where it reduces the average ASR from PAD’s $18.65\%$ to $8.07\%$. 

Furthermore, both PAD and PAD-FT demonstrate robustness and consistency across different poison rates. For example, in CIFAR-100 with WaNet attack, PAD-FT and PAD maintain an ASR below $0.7\%$ even at a $10\%$ poison rate, significantly outperforming DBD with an ASR of $97.19\%$ at $\rho=10\%$. 

These results highlight that PAD and PAD-FT provide reliable, adaptive defenses with strong attack mitigation across diverse scenarios.

\subsection{Ablation Study}
An ablation study is conducted to evaluate how $N_c$, the size of the self-purified dataset $\mathcal{D}_c$ influences the effectiveness of the proposed PAD-FT mechanism. 

\begin{table}[h!]
    \centering
    \caption{ACC (\%) and ASR (\%) of PAD-FT with different $N_c$ on CIFAR-10 attacked by BadNets with $\rho=10\%$}
    \begin{tabular}{c c  c c c c c c c }
    \toprule 
    $N_c$ & 1000  & 2000  & 3000  & 4000 & 5000 & 6000\\
\midrule
ACC & 83.43 & 84.70 & 83.91 & 82.02 & 82.38 & 82.22 \\
 ASR & 16.10 & 9.96 & 8.82 & 10.96 & 13.11 & 13.64\\
    \bottomrule
    
    \end{tabular}
    \label{tab:ablation}
\end{table}

Table \ref{tab:ablation} demonstrates that ACC initially increases as $N_c$ grows, benefiting from more purified data used for fine-tuning. However, as $N_c$ continues to increase, ACC starts to decline, likely because more poisoned data is being selected into $\mathcal{D}_c$. Similarly, ASR decreases at first but rises again as more poisoned data are included. This illustrates a clear tradeoff between increasing the size of the purified data  and the model performance in terms of ACC and ASR. 

\section{Conclusion}
This paper proposed a novel lightweight post-training backdoor defense mechanism PAD-FT. By introducing a new data purification method, PAD-FT effectively disinfects poisoned models without requiring additional clean data. The classifier-only fine-tuning in PAD-FT highlights its lightweight nature, making it easy to implement. Extensive experiments demonstrate the effectiveness and superiority of PAD-FT across a variety of datasets and backdoor attack scenarios. Therefore, our PAD-FT mechanism offers a practical and efficient solution to the challenge of backdoor attacks.

\bibliographystyle{IEEEbib}
\bibliography{refs}

\end{document}